# Imputing missing not-at-random longitudinal marker values in time-to-event analysis: fully conditional specification multiple imputation in joint modeling

Murad Havi*[1], Agay Nirit[1,2], Dankner Rachel[2,3]

[1] Biostatistics and Biomathematics Unit, Gertner Institute, Sheba Medical Center,
    Tel-Hashomer, Israel

[2] Public Health Unit, Gertner Institute, Sheba Medical Center,
    Tel-Hashomer, Israel

[3] Department of Epidemiology and Preventive Medicine, Sackler Faculty of Medicine, School of Public Health, Tel-Aviv University, Tel-Aviv, Israel

*CORRESPONDING AUTHOR: Havi Murad (PhD), Biostatistics and Biomathematics Unit, Gertner Institute of Epidemiology & Health Research Policy, Sheba Medical Center, Israel. HaviM@gertner.health.gov.il ; Havimurad@gmail.com

Abstract

We propose a procedure for imputing missing values of time-dependent covariates in a survival model using fully conditional specification. Specifically, we focus on imputing missing values of a longitudinal marker in joint modeling of the marker and time-to-event data, but the procedure can be easily applied to a time-varying covariate survival model as well. First, missing marker values are imputed via fully conditional specification multiple imputation, and then joint modeling is applied for estimating the association between the marker and the event. This procedure is recommended since in joint modeling marker measurements that are missing not-at-random can lead to bias (e.g. when patients with higher marker values tend to miss visits). Specifically, in cohort studies such a bias can occur since patients for whom all marker measurements during follow-up are missing are excluded from the analysis. Our procedure enables to include these patients by imputing their missing values using a modified version of fully conditional specification multiple imputation. The imputation model includes a special indicator for the subgroup with missing marker values during follow-up, and can be easily implemented in various software: R, SAS, Stata etc. Using simulations we show that the proposed procedure performs better than standard joint modeling in the missing not-at-random scenario with respect to bias, coverage and Type I error rate of the test, and as good as standard joint modeling in the completely missing at random scenario. Finally we apply the procedure on real data on glucose control and cancer in diabetic patients.

Keywords: joint modeling, MICE imputation, fully conditional specification imputation, multiple imputation, missing data, partially observed time varying covariate

1. Introduction

When relating a series of repeated measurements, e.g. a marker like blood glucose, to a subsequent event, e.g. cancer, there are two possible approaches: (*i*) A Cox model with the marker as a time-varying covariate (TVC) [1,2] and (*ii*) The Joint Modeling (JM) framework for longitudinal and time-to-event data (JM) [2,3,4,5]. The first approach, very common in use, estimates the association between the observed marker values and the event, assuming that marker values are fixed between visits and measured without error. The second approach estimates the association between the predicted continuous latent marker values and the event, and requires further assumptions on the form of the latent marker and on the baseline hazard. When there are missing values in the marker, the JM approach is known to perform well as long as missing is at random (MAR) [2]. In observational data, marker measurements are sometimes missing to the extent that some people in the cohort have no measurements at all during follow-up period. In the JM approach, these individuals would be excluded from the analysis. However, if there is a tendency of the higher (or lower) measurements to be missing (i.e. "not missing at random" - NMAR) then excluding this "all-missing" subgroup of people may lead to biased results.

In this paper we propose a method which overcomes this problem and yields unbiased results: a two-step procedure where in the first step the missing marker values are imputed via a modified version of Fully Conditional Specification (FCS) multiple imputation (MI), and in the second step Joint Modelling (JM) is applied to each imputed data set. A pooled estimate is then obtained by Rubin's rule [6]. The main advantage of this procedure is that unlike the standard JM, it enables to include the subgroup with no measurements at all (the "all-missing" subgroup). Such a subgroup may exist in every cohort study with a longitudinal marker. To successfully impute the missing observed values in this subgroup, we propose to include an indicator of belonging to it, an idea that was not suggested previously. In addition, we suggest using available measurements that occurred after the event, in order to improve the imputation; this idea will be further discussed in Section 5.

In a previous work we developed a time-sequential MI [7] which is an approximation of the FCS (conditioning only on the marker past values in the imputation model) and we showed that it performed well, even under NMAR. In this paper we focus on the FCS MI approach, which is more accurate since it is doubly iterative (within time and among times; [8]). In general, FCS MI specifies a series of univariate models for the conditional distribution of each partially observed variable given the other variables. Each partially observed variable is treated as an independent variable, so that in the case of a partially observed TVC (marker), the timely order of repeated measurements is ignored. This means conditioning on past values as well as on future ones [9]. The imputation step can be applied using the MI procedure in SAS with FCS statement or using similar packages in other software, e.g. the 'mice' package in R or 'ice' package in Stata.

A researcher may be interested in estimating the association of the time-to-event with observed marker values, (e.g. for prediction purposes) or with the true latent marker values (after correction for measurement error). While in our previous work mentioned above, we were interested in the first goal (therefore used at the second step a time-varying Cox model with discrete times), here we are interested in the association of the latent true marker values with survival, therefore using at the second step JM with continuous time. For the JM step we used 'JM' package in R [10], which is very common in use. Note that our suggestion for the imputation at the first step is also relevant for a second stage of time-varying Cox model.

We focus on a lag1 Cox model, i.e. association of time-to-event with previous marker value, but our method can be easily generalized to different lags or to a current value model, when time-to-event is associated with the current marker value. The functional form of the marker over time in the substantive time-to-event model (e.g. current value,

lag1, lag2 etc.) influences the imputation model, since the imputation model should include not only the other covariates in the substantive model [11,12] but also the outcome variable. This is important as to maintain the association between the covariate and the outcome [13,14]. The appropriate way to incorporate the outcome variable in the imputation model for the marker depends on the type and specification of the substantive model [9]. Incompatibility between the imputation model and the substantive model may cause bias. In this paper we define the correct form of the outcome for various substantive time-to-event models.

Using simulations based on real data, we compared the performance of the standard JM with the suggested two-step procedure of FCS MI followed by JM. We examined two versions of the imputation model: (*i*) a standard version of FCS MI; (*ii*) a modified version of FCS MI, including an indicator for the "all-missing" subgroup. We examined two types of missingness: completely missing at random (CMAR), and not missing at random (NMAR) in two levels - weak and strong.

The paper is structured as follows. First we describe the more frequently adopted standard JM contrasting it with our two-step procedure. Next, we describe how to perform FCS MI for a time-varying covariate, including past values as well as future values of the marker in the imputation model, as well as a function of the outcome variable. We focus on the lag1 Cox model, but provide the correct function of the outcome to be included in the imputation model also for other time-to-event models (current value and different lags). We then present simulations generated under the alternative hypothesis and under the null hypothesis assuming different types of missing data, and compare the results of our procedures to standard JM. Next, we present an application of our procedure to real data of glucose control variables among diabetic patients from a large Israeli cohort. Recommendations regarding how to identify the missing pattern (type, direction and level) in real data are provided. Finally, we discuss the performance of our procedure and its robustness to the degree of, its advantages and limitations.

## 2. Methods

### 2.1 Standard JM (lag1 model, [2]):

**Step 1:** Assume we know the true and unobserved value of the marker at time

*t-1*, $m_i(t-1)$, then

$$\lambda\big((t|\, M_i(t-1))\big) = \lambda_0(t-1)\, exp\{\beta_1 female + \beta_2 older + \alpha m_i(t-1)\} \quad (1)$$

**where**

$M_i(t-1) = \{\, m_i(s), 0 \leq s < t-1\,\}$ **is the longitudinal history**

**Step 2:** From the observed longitudinal response $W_i(t)$ we reconstruct the covariate history for each subject using a Mixed Effect Model:

$$W_i(t) = m_i(t) + \varepsilon_i(t) = \underline{x}_i^T(t)\underline{\beta} + \underline{z}_i^T(t)\underline{b}_i + \varepsilon_i(t)\,; \quad (2)$$

where  $\varepsilon_i \sim N(0, \sigma^2)\,;\quad \underline{b}_i \sim N(0, D)$

**Step 3:** The two processes are easily associated as a product, since in the shared parameters framework they are independent conditional on random effects. Estimation is done by maximum likelihood [2].

***Figure 1***

### 2.2 Multiple Imputation

#### 2.2.1 Bayesian MI [9,15]

In Bayesian parametric MI, to multiply impute missing values in *W* we specify a parametric model $f(W|Z, Y, \theta)$, $\theta \in \Omega$ for the conditional distribution $f(W|Y, Z)$. To create the $m^{th}$ imputed dataset we first draw θ(*m*) from its posterior distribution given the observed data $\{(Y, Wobs, Z); i = 1, \ldots, n\}$ and a (usually noninformative) prior $f(\theta)$. For each subject the missing values (if any) $W^{mis}$ are imputed by taking a draw from the density $f(Wmis|Wobs, Y, Z, \theta(m))$ implied by $f(W|Y, Z, \theta(m))$.

### 2.2.2 Standard FCS MI algorithm ([9] - an approximation to Bayesian MI)

For each partially observed covariate $W_j$, where $j=1, …, p$, we posit an imputation model $f(W_j | W_{-j}, Z, Y, \theta_j)$, with parameter $\theta_j$, where

$W_{-j} = (W_1, …, W_{j-1}, W_{j+1}, ….W_p)$. In our case it is a linear model since our $W$, log transformed HbA1c, is continuous.

Let $w_j^{obs}$ and $w_j^{mis}$ denote the vectors of observed and missing values in $W_j$ for the $n$ subjects.

Let $w_j^{mis(k)}$ denote the imputations of the missing values $w_j^{mis}$ at iteration $k$ and let $w_j^{(k)}=(w_j^{obs}, w_j^{mis(k)})$ denote the vector of observed and imputed values at iteration $k$.

Let $w_{-j}^{(k)}=(w_1^{(k)},…, w_{j-1}^{(k)}, w_{j+1}^{(k-1)},…, w_p^{(k-1)})$

**Step 0**

Replacing missing values in each $w_j$ with randomly selected observed values from the same variable.

Then iterative algorithm begins.

**Iteration $k$**

We draw $\theta_j$ from its posterior distribution given $w_j^{obs}$, $w_{-j}^{(k)}$, $z$, $y$.

This is equal (up to a constant of proportionality) to the product of the prior $f(\theta_j)$ and the likelihood corresponding to fitting the imputation model for $W_j$ to subjects for whom $W_j$ is observed, using the observed and most recently imputed values of $W_{-j}$.

Missing values in $W_j$ are then imputed from the imputation model using the parameter value drawn in the preceding step.

After a sufficient number of iterations it is assumed that the algorithm has converged to a stationary distribution, and the final draws of the missing data form a single imputed dataset. The process is then repeated to create as many imputed datasets as desired.

**An example for Iteration $k$ for our lag1 substantive model with $j=1,2,3$:**

Let $i=1, \ldots, n$, an index for the subject.

Let $Y_{ij}$ be an indicator for occurrence of the event at time-period $j$.

Imputation models for subject $i$:

$$ln(w_{i1}) = \theta_{0_1} + \theta_{1_1} z_i + \theta_{2_1} y_{i2} + \theta_{3_1} ln(w_{i2}^{(k-1)}) + \theta_{4_1} ln(w_{i3}^{(k-1)}) + \xi_{1i};$$

$$ln(w_{i2}) = \theta_{0_2} + \theta_{1_2} z_i + \theta_{2_2} y_{i3} + \theta_{3_2} ln(w_{i1}^{(k)}) + \theta_{4_2} ln(w_{i3}^{(k-1)}) + \xi_{2i};$$

$$ln(w_{i3}) = \theta_{0_3} + \theta_{1_3} z_i + \theta_{2_3} y_{i4} + \theta_{3_3} ln(w_{i2}^{(k-1)}) + \theta_{4_3} ln(w_{i3}^{(k-1)}) + \xi_{3i};$$

$$\xi_1 \sim N(0, \sigma_{\xi_1}^2) \; ; \; \xi_2 \sim N(0, \sigma_{\xi_2}^2) \; ; \; \xi_3 \sim N(0, \sigma_{\xi_3}^2)$$

Note that this procedure imputes missing values for all time-periods, including those occurring subsequent to the event (i.e. after end of follow-up). However, these imputed values will be deleted prior to the time-to-event analysis.

2.2.3 The modified FCS MI algorithm

When missing is NMAR, the "all-missing" subgroup tends to have higher or lower marker values than the rest of the cohort. The direction of NMAR is determined by whether those with higher values tend to miss more visits, or the opposite. Therefore, we suggest to modify the algorithm described in 2.2.2 and add to the imputation model a special indicator for belonging to the "all-missing" subgroup. This indicator will improve the imputation in this subgroup when data are NMAR. It can also help the researcher understand the type of missing data, and its direction. Detailed explanation will be given in the results section.

2.2.4 Correct form of the outcome in the imputation model

As already mentioned, it is important to include a correct form of the outcome in the imputation model as to make it substantive model compatible (Bartlett, 2015). The correct form of the outcome variable used in the imputation model depends on the functional form over time of the partially observed marker used in the time-to-event model. For example, in our previous paper (Murad et al, 2019), the TVC (i.e. longitudinal marker) in the substantive model was the average *HbA1c* level over the previous four time periods. We therefore note that *HbA1c* for a specific time-period $j$ impacts on $\lambda(j)$ only when $t = j + 1, j + 2, j + 3, j + 4$, and only for patients who are still in the risk set for cancer at $t = j + 1$. In appendix A of that paper, we showed that for the imputation procedure to be approximately valid in this case, it is sufficient to include in the imputation model an indicator for an event in one of the subsequent four time-periods, together with the cumulative baseline hazard for the subsequent time-periods that the patient remains in the risk setup, to a maximum of four periods. In this paper, the substantive model is a lag1 time-to-event model, which is a special case of the model from the previous paper. We therefore note that *HbA1c* for a specific time-period $j$ impacts $\lambda(j)$ only when $t = j + 1$, i.e., the subsequent time-period, and only for patients who are still in the risk set for cancer at $t = j + 1$. Therefore, for the imputation procedure to be approximately valid, it is sufficient to include in the imputation model an indicator for the event in the subsequent time-period. In this special case, the cumulative baseline hazard is taken over a single time-period (the subsequent one), and therefore it can be treated as a constant for all subjects. Note that because we use lag1 time-to-event model, if a subject is not in the risk set at a specific time-period $j+1$, we do not use his marker value at time-period $j$.

Similarly, this function of outcome can be generalized to any number of previous periods back. For a current value model (where the current value of the marker is associated with the outcome), it is sufficient to include an event indicator for the current time-period.

2.2.5 The two-step procedure

In the suggested procedure we first impute the missing marker values using FCS MI and then apply JM on each completed data set (both non-missing and imputed values), and use Rubin's rule for pooling of results. In this way, the "all-missing" subgroup is not excluded from the analysis.

## 3. Simulations

### 3.1 Simulations under $H_1$

We conducted simulations based on the prevalent diabetes group data from the Clalit HMO, where the event was cancer. We generated 7 time-periods of the TVC *HbA1c*. The confounders Z were as follows: age group (>70, 50–70 years) and sex, generated from Bernoulli distribution with probabilities: 0.47 and 0.5, respectively.

To be able to use lag1 Cox model, HbA1c levels were generated for eight time-periods (i.e. for one additional time-period) from a linear mixed model with random intercept ($a_i$) and random slope ($b_i$), based on Clalit HMO data:

$$ln(HbA1c_i(j)) = m_i(j) + \varepsilon_i(j)$$
$$= \alpha + a_i + \beta_1 j + b_i j + \beta_2 female + \beta_3 age\_older + \varepsilon_i(j) ;$$
$$a_i \sim N(0, \sigma^2_a) ; b_i \sim N(0, \sigma^2_b) ; \varepsilon_i \sim N(0, \sigma^2_\varepsilon)$$
$$Where\ \alpha = 2.04 ; \beta_1 = -0.02 ; \beta_2 = 0.02 ; \beta_3 = -0.07 ;$$
$$\sigma^2_a = 0.0236 ; \sigma^2_b = 0.0003 ; \sigma^2_\varepsilon = 0.006 \Leftrightarrow \rho_{intracluster} \cong 0.8$$

The cancer event (y/n) for each time-period was generated by a logistic discrete-time survival model [16] with the covariates: age group, sex and log-scaled true (i.e. latent) value of previous HbA1c ($m_i(j-1)$). The respective coefficients for these covariates were as follows: 0.69 (HR=2.0, older vs. younger), -0.3 (HR=0.74, females vs. males), 1.4 (HR=1.15 for a 10% increase in HbA1c). The intercept was fixed on -5, to represent a baseline hazard of 0.0067 for each time-period. For simplicity, death or other censoring, i.e. aging, were not introduced. Then, for compatibility with JM, follow-up time was transformed from discrete to continuous using exponential distribution. For each subject, random intercept and slope as well as a random residual were drawn from a normal distribution with zero mean and variances of 0.0236, 0.0003 and 0.006 respectively.

**Generation of Missing data**

We generated 400 simulations of 4000 subjects, each under different type of missing. We first created fully-observed data of HbA1c levels at each time-period. Then, we used the following shared parameter model for generating informative missing data [17]:

$$logit(p(HbA1c_{ij} = miss)) = -0.405 + \gamma_1 * a_i + \gamma_2 * b_i \qquad (3)$$

In this model, when $\gamma_1 = \gamma_2 = 0$, then missing values are completely missing at random (CMAR or non-informative missing) and $p(HbA1c_{ij} = miss) = 0.4$, i.e. 40% of values are missing at each time-point. The largest $\gamma_1$ and $\gamma_2$ are, the more informative the missing is (the stronger the degree of NMAR). We chose two levels of NMAR: (*i*) weak: $\gamma_1 = 2; \gamma_2 = 5$ (the mean of missing HbA1c values, before introducing missing values in the simulated data, was 6% higher than mean of non-missing values); and (*ii*) strong: $\gamma_1 = 20; \gamma_2 = 25$ (the mean of missing HbA1c values was 26% higher than the mean of non-missing values); The overall missing percent was 40% under NMAR as well.

We first calculated the probability for a missing HbA1c value for each subject and for each time-period from equation (3) under different missing types, and then we randomly generated an indicator for missing value from Bernoulli distribution, using this probability.

**Analysis**

After creating data sets with missing data, we applied the standard JM. Then we applied the two versions of FCS (standard and modified) where we first imputed each missing data five times and then applied JM to each of the completed data sets and combined the results using Rubin's rule. As a gold standard, we also applied JM to the fully-observed simulations, before the creation of missing values.

For each method, we averaged the regression coefficients of the marker over the 400 simulations, and computed the empirical variance. In addition, Percent Bias (PB), root mean squared error (RMSE) and coverage were calculated, where for FCS MI the confidence intervals were based on a *t*-distribution with appropriate degrees of freedom (Schaffer, 1997). For the FCS MI procedures, the proportion of variance due to missingness was calculated and averaged over the 400 simulations.

To compare the two versions of FCS MI (standard and modified) with regard to accuracy of imputed values, we compared fully-observed values to imputed values in each time-period (averaged over 5 multiples).

### 3.2 Simulations under $H_0$

It is of interest to investigate the Type I error rate of the test in each method under different missing types, since occasionally a method can yield a biased estimate for a parameter while the corresponding statistical test of the null hypothesis for that parameter is valid. Therefore, a second set of simulations was generated under the null hypothesis, in a similar way to the description in 3.1, but where the association parameter in the JM, $\alpha$, equals zero (see equation (1)). To estimate the type I error probability of the test for $\alpha = 0$, we increased the number of simulations to 1600. Two types of missing data were generated, weak and strong NMAR. The type I error probability was estimated as the percentage of rejections of $H_0: \alpha = 0$ over the 1600 simulations (in FCS MI methods after the Rubin's pooling phase), using the nominal two-sided 5 percent level, that is $|t| \geq |q_{t_{0.025, df}}|$ in each simulation, using the t statistic, $(\hat{\alpha} - 0)/(Model\_SE /\sqrt{1600})$, where $Model\_SE$ is the model-based standard error from each simulation. Note that *df* is calculated for each simulation [12].

### 3.3 Results of Simulations

Figure 2 presents observed marker values (*W*) of 10 randomly selected subjects. It indicates variability in intercepts as well as in slopes.

***Figure 2***

Table 1 presents the results of the 400 simulations. The estimator for log-HR of the association of HbA1c and time to cancer when using fully-observed values of HbA1c (i.e. before inducing missing values) was slightly biased towards the null (true value 1.4, mean estimated value 1.31, PB = - 7%), presumably due to the finite sample size.

In the CMAR scenario, the standard JM estimator (excluding the "all-missing" subgroup) was only slightly biased downward (1.31, PB = -6%) and very similar to the fully-observed estimator. Standard FCS MI and modified FCS MI yielded similar estimates (1.27 and 1.28 respectively, PB = -9%). In all three methods, the coverage was slightly lower than the nominal level of 95%.

In the weak NMAR scenario, the standard JM yielded a biased downward estimator (1.19, PB = -15%) and a lower than expected coverage (74%), whereas standard FCS and modified FCS yielded estimates close to the fully-observed one (1.27 and 1.29 respectively) and a better coverage (90%).

In the strong NMAR scenario, the standard JM yielded a very biased estimator (0.94, PB = -33%) and a very low coverage (43%). Standard FCS MI performed a bit better in terms of bias (1.03, PB = -26%) but maintained a too low coverage (58%). Modified FCS MI was the best among the three methods yielding an estimate close to the fully-observed (1.29, PB = -8%) and a coverage of 95%.

***Table 1***

Overall, the results show that the stronger the NMAR level, the larger the difference between the estimates of the two versions of FCS. Thus, comparing results of the two-step procedures using the two versions of FCS can indicate the degree of NMAR in the data.

Another measure of degree of NMAR is the proportion of variance due to missingness ($\lambda$). In our simulations it increased with the degree of NMAR, and overall it was similar in the standard and the modified FCS. However, in the strong NMAR, $\lambda$ was slightly higher in the modified FCS compared to the standard FCS, meaning a higher heterogeneity in the five estimates obtained from the five completed data sets. In order to compare performance of the two FCS imputation models (standard and modified), we compared the fully-observed values to the completed values (both non-missing and imputed values) of the marker in each time-period. Table 2 presents the fully-observed (W) and completed values (following imputation) of the log-transformed marker in each time-period, averaged over the five data sets, in the "all-missing group" and in the rest of the sample separately. The table shows that in the rest of the sample, the fully-observed values are very similar to the completed values in all missingness scenarios. In the "all-missing" subgroup, the stronger the NMAR level was the higher the fully-observed values were compared to completed values. However, the modified version of FCS imputation yielded closer values to the fully-observed than the standard version, even in the strong NMAR, implying that including the special indicator improves the imputation within the "all-missing" subgroup.

***Table 2***

Furthermore, comparison of the imputed values yielded by the two versions of FCS in the "all-missing" subgroup can indicate even more clearly the type of missingness in the data and its direction. In our simulations, the stronger the NMAR was, the higher the values yielded by the modified version relative to the standard version were, and the further from 1 was the ratio between the mean of completed values (both non-missing and imputed values) yielded by the two versions. A higher than 1 ratio implies that the special indicator included in the modified version increased the imputed values, indicating that missing values were more probable among higher values of HbA1c. For example, in the strong NMAR scenario the ratio was 1.05. This is consistent with the way the missing values were generated (see Section 3.1). Such a comparison between the versions is also feasible in real data, for assessing the degree of NMAR and its direction, as will be demonstrated in Section 4.

Regarding type I error rates (Table 3), in the weak NMAR scenario all methods performed well, yielding rates close to the nominal level of 5%, with the standard JM test performing slightly worse than the FCS-JM procedures (5.9% vs. 5% and 4.8%). In the strong NMAR scenario, the standard JM test yielded a higher than expected Type I error rate (9.2%) whereas FCS-JM tests yielded slightly lower than expected rates (3.9% and 3.1% respectively for standard and modified FCS MI).

**\*\*\*Table 3\*\*\***

## 4. A motivating example

We used data from a population-based historical cohort study described in details in Dankner et al. [18]. The cohort includes all subjects aged 21–89 years from Clalit HMO, the largest HMO in Israel, followed between 1 January 2002 and 31 December 2012. Time axis was divided into 6-month intervals. The data file includes detailed high-quality demographic, clinical, and pharmaceutical information and was linked to the Israel National Cancer Registry for cancer morbidity.

The clinical aim in this example was to evaluate the association between HbA1c level in the previous 6-months period, treated as a time-dependent exposure, and the risk of pancreatic cancer, among patients with diabetes. The data file included approximately 153,000 diabetic patients, all diagnosed with diabetes in the years 2002-2012 (incident cases). Follow up for cancer started two years after diabetes diagnosis. We first identified the subgroup with no HbA1c measurements at all during their follow-up time (n=33,420). The role of this subgroup was described in section 2.2.3. Table 4 presents baseline characteristics of the study population, by the "all-missing" and the rest of the sample. As can be seen, the "all-missing" subgroup is slightly younger than the rest of the sample, their mortality is almost doubled (17% vs. 8%) and their follow-up time is shorter (3.6 vs. 5 years). This might indicate that the patients in this subgroup are in a poorer health condition which prevented them from complying to lab visits, in which HbA1c level is measured.

***Table 4***

Due to the substantial amount of missing data in observed HbA1c (40% in any specific time-period), we applied the proposed FCS MI in two versions, as described in 2.2.2-2.2.3, with ten imputations per missing value [19]. Next, according to our proposed procedure, we applied JM on each completed data file and used Rubin's rule to pool the results. These results were compared to standard JM results (i.e. without the imputation step, excluding the "all-missing" subgroup). Observations at death or age 90 were censored in the JM model for cancer, yielding estimates of cause-specific hazard ratios. Hba1c values were log transformed.

Table 5 presents the hazard ratios for a 1 unit increase in log-HbA1c values for the three methods. The "all-missing" subgroup is omitted in the standard JM, so that the model estimates the association within the rest of the sample ($n$=119,383). This yielded a hazard ratio of 3.61, which corresponds to a statistically significant 41% increase in the hazard of pancreatic cancer for a 10 percent increase in HbA1c. When applying our two-step procedure to the whole sample ($n$=152,803), lower hazard ratios were obtained (3.41 and 3.35 for standard and modified FCS versions respectively). As already shown in Section 3.3, the difference between the results of the two FCS versions (3.41 vs. 3.35) indicates that the missingness in the data is NMAR.

***Table 5***

We further checked this hypothesis by comparing the imputed values of the two FCS versions. Table 6 presents the completed values (both non-missing and imputed values) yielded by the two versions of FCS (standard versus modified FCS) in each time-period, for the "all-missing" subgroup and for the rest of the sample separately. As expected, within the rest of the sample both versions produce similar values. Within the "all-missing" subgroup, the table shows that the imputed values yielded by the modified FCS version are slightly lower than those yielded by the standard FCS version, implying NMAR: lower marker values tend to be missing (an opposite direction to the simulations). This means that patients who are more glucose-balanced are those who tend to miss visits. In order to assess the degree of NMAR we calculated the ratio of completed values (both non-missing and imputed values) in the modified versus the standard version of FCS. The average ratio over all time periods was 0.97. To make it comparable to the ratio in the simulations, where direction of NMAR was opposite, we looked at 1/ratio, which is 1.04, very close to the strong NMAR simulations. The proportion of variance due to missingness ($\lambda$, Table 4) also supports NMAR since it is not close to zero (0.2 and 0.26 for the standard and modified FCS versions respectively).

***Table 6***

5. Discussion

We focus on associating time-to-event with a partially observed longitudinal marker (time varying covariate). In a previous paper we proposed to impute the missing values in the observed marker time-sequentially using chained equations MI (an approximation for FCS), while the imputation model takes into account the outcome. Then, we used a discrete-time Cox model with time-dependent covariates to estimate the association between the observed marker values and time-to-event. We showed that this method performs well even when missing is not at random [7]. However, this procedure does not correct for measurement error in the marker values, which might attenuate the association of interest. In this paper we continue this line of research, using a more accurate imputation method (FCS) and joint modelling for continuous follow-up time instead of a discrete Cox model with a time-varying covariate. The advantage of joint modelling is that it estimates the association of the true marker values with time-to-event, i.e. after correction for measurement error. A researcher may be interested in estimating the association with the observed marker values, (e.g. for prediction purposes) or with the true latent marker values.

In the standard JM framework for longitudinal and time-to-event data, only available marker measurements are used in the estimation of the longitudinal process, and subjects with no marker measurements during follow-up ("all-missing" subgroup) are excluded. When measurements are missing at random this performs well. However, when missing is not at random, missingness depends on the unavailable marker values, thus this "all-missing" subgroup might represent a distinct profile of patients, and excluding it might severely bias the association, especially when its size is considerably large. Therefore, in this paper, we present an improved version of standard shared parameters JM to estimate the association of a longitudinal marker and the time-to-event, where we impute the missing values in the marker via FCS MI before applying the JM. Optionally, when missing data is expected to be NMAR, it is possible to jointly model the NMAR as an additional process within the joint modeling framework; however, this may be complicated, and will not prevent exclusion of the "all-missing" subgroup.

Our imputation model includes past values as well as future values of the marker and additionally an appropriate form of the outcome (substansive-model compatible; Bartlett, 2015). It should be emphasized that our general idea of FCS imputation could be applied in any software (e.g. SAS: MI procedure with fcs statement, R: 'mice' package or 'smcfcs' package, Stata: 'ice' package etc.), with the outcome included in the imputation model in a form that is compatible with the substantive survival model used. Having obtained complete data sets of the longitudinal process (i.e. both non-missing and imputed values of the observed marker), one can apply any time-to-event model on each of them and pool the results using Rubin's rule. We chose the Shared parameters joint modelling and used 'JM' package in R [10], since it is very common in use and also corrects for measurement error in the marker.

To our knowledge, this paper is among the first studies applying FCS MI to a partially observed marker (TVC) associated with time-to-event outcome. Previous studies demonstrating the use of FCS in imputing covariates in survival analysis, involved fixed in time covariates only [20,21,22]. For time-varying covariates such as markers, the question is what form of the outcome should be included in the imputation model, to make it compatible with the outcome model. In a previous paper we provided a suggestion accompanied by a proof for the case of substantive survival models with exposure in the previous four time-periods [7]. We suggested using an indicator for event in the four subsequent time-periods together with the cumulative baseline hazard for the subsequent time-periods that the patient remains in the risk setup, to a maximum of four periods, for the imputation procedure to be approximately valid. This suggestion can be generalized to substantive Cox models with any number of periods back (lag) in the exposure/marker.

Others who studied TVCs in a time-to-event model investigated only missing at random type of data. For example, Bhattacharjee [23] dealt with missing values in the outcome as well as in the longitudinal marker. He used four single-imputation methods for the marker in shared parameters JM, but did not use FCS MI. Welch [24] used the two-fold FCS MI proposed by Navailanen [8] in a Cox model associating the marker value in the first time-period with time-to-event. Betancure [25] dealt with joint modelling of multiple partially observed markers and survival and proposed a method to impute the missing marker values at the event times only, under MAR, using chained equations (but not doubly iterative FCS). Bartlett [9] presented FCS MI for TVC but did not associate it with survival.

In this paper, we additionally present a novel idea for improving the imputation within the "all-missing" subgroup, i.e. those with all marker values missing during follow up, even when data is not missing at random. We suggest including an indicator for this subgroup in the imputation model, hence called the modified version of FCS MI. Also, for imputing missing marker values within this subgroup, we recommend using observed marker measurements that occurred after the event. Indeed, these measurements might be affected by the event, but still, they can be useful in imputing missing values, because they are correlated with pre-event values at least to some extent. Therefore using them further improves the imputation, as also shown by our simulations (see Table 5). Hence this modified version is especially relevant when the longitudinal marker can be observed after the event has occurred (i.e. when the event of interest is not death).

We examined the performance of our suggested procedures versus standard JM using simulations and showed that when missing is not at random, the estimate obtained by our procedures is less biased than that of the standard JM, and the type-1 error rate is closer to the nominal level. The results are even better with the modified FCS version, most probably since the imputation within the "all-missing" subgroup is improved (see Table 5). In view of its similarity to standard multiple imputation, the suggested imputation procedure is expected to perform well when the MAR assumption holds true, just as standard JM without imputation.

In real data, unlike in simulations, one cannot distinguish between MAR and NMAR types of data, since fully-observed data are unavailable. The proportion of variance due to missingness, $\lambda$, a parameter which is very common in use, might indicate the type of missingness since the closer it is to zero, the closer the missingness to MAR. However, $\lambda$ does not provide the direction of missingness, i.e. do missing values tend to appear in higher or lower values of the marker. In cases where the "all-missing" subgroup is large, we suggest to evaluate the type of missingness by comparing the completed values (after imputation) obtained by the two versions of FCS (standard and modified) within the "all-missing" subgroup. This can give us a hint about the direction and degree of NMAR in our data: the similar the values yielded by the two versions, the less the data is expected to be NMAR. Indeed, in our application to real data we found that the completed values in the modified FCS version were slightly lower than in the standard FCS version, implying NMAR: lower marker values tend to be missing (in opposite direction to the simulations). Calculation of the average of the ratio of the completed values over all time-periods revealed quite a strong degree of NMAR (see end of Section 4).

Among the limitations of this study are that JM might be severely biased when the baseline hazard or the shape of the longitudinal trajectories are misspecified. In addition, our results and conclusions are limited to the shared parameters missing data mechanism model, which is appropriate for the shared parameters JM.

In future research we would like to examine JM robustness to the mixed model under which the simulations were created, and generalize this work to the case of several markers with missing data.

Acknowledgements

We thank Prof Laurence Freedman from the Biostatistics and Biomathematics Unit in the Gertner Institute (Israel) for his support, encouragement and intuitions.

## 6. Software

```
******************************************************************;
 *SAS PROGRAM - FULLY CONDITIONAL SPECIFICATION IMPUTATION;
  *PROGRAMMERS: HAVI MURAD (PhD), BIOSTATISTICS AND BIOMATHEMATICS
  UNIT, GERTNER INSTITUTE;
 *DATE: MARCH 2023;
 *CREATING 1 SIMULATION of 4,000 SUBJECTS WITH CMAR MISSING DATA IN
  HbA1c (40%)AT EACH OF 7 TIME-POINTS.
 *EVENT IS CANCER AND WE ASSUME NO DEATH OR LOST TO FOLLOW-UP;
 *THIS PROGRAM APLLIES FCS imputation FOR ln(HbA1c) USING PAST AS
 WELL AS FUTURE VALUES AND THE CORRECT FORM OF THE OUTCOME;
 *THEN JOINT MODEL is applied to each completed data set (after
  imputation)'JM' package in R (not shown);
 *Finall RUBIN'S RULE is applied for pooling the results (not
  shown);

*ntime is continuous follow-up time until cancer or end of follow-
up;

*ntime_disc is discrete follow-up time;

*H1-H7 is observed longitudinal marker values (the W's) ;

*ny1-ny7 are indicators for cancer event at each time-point;

*D1-D6 indicators for cancer event in subsequent time-point;

*omit is an indicator for the "all-missing" subgroup;

******************************************************************;

*******************************************************************;
*IDENTIFICATION OF THE "ALL-MISSING" SUBGROUP: DEFINING THE "OMIT"
INDICATOR;
******************************************************************;

data b;
 set havif;
 array hb{*} nH1-nH7;
  miss=0; nmeas=0;
 do i=1 to 7;
  time=i-0.5;
   if  time<=ntime then do;
        if hb[i]=. then miss=miss+1;
         nmeas=nmeas+1;
```

```
        end;
    end;
    if miss=nmeas then omit=1; else omit=0;
     if ntime<1 then delete;
      *drop time;
    run;

*********************;
*IMPUTATION STARTS HERE;
*********************;
```

data havib;
  set b;
ntime_disc=ceil(ntime);
lH1=log(nH1);
lH2=log(nH2);
lH3=log(nH3);
lH4=log(nH4);
lH5=log(nH5);
lH6=log(nH6);
lH7=log(nH7);
array ny {*} ny1-ny7;
array D {*} D1-D6;
do i=1 to 7;
  if i<=6 then do;
    if ny[i+1]=1 then D[i]=1; else D[i]=0;
  end;

```
/*modified FCS MI*/
proc mi data=havib seed=24072023 nimpute=5 out=mult_fcs_allomitfin_CMAR;
   class female old;
      FCS nbiter=10 reg (lh1 = lh2 lh3 lh4 lh5 lh6 lh7 female old D1 /details);
      FCS nbiter=10 reg (lh2 = lh1 lh3 lh4 lh5 lh6 lh7 D2 female old omit/details);
      FCS nbiter=10 reg (lh3 = lh1 lh2 lh4 lh5 lh6 lh7 D3 female old omit/details);
      FCS nbiter=10 reg (lh4 = lh1 lh2 lh3 lh5 lh6 lh7 D4 female old omit/details);
      FCS nbiter=10 reg (lh5 = lh1 lh2 lh3 lh4 lh6 lh7 D5 female old omit/details);
      FCS nbiter=10 reg (lh6 = lh1 lh2 lh3 lh4 lh5 lh7 D6 female old omit/details);
      FCS nbiter=10 reg (lh7 = lh1 lh2 lh3 lh4 lh5 lh6 female old omit/details);
   var lH1 lH2 lH3 lH4 lH5 lH6 lH7 D1-D6 female old omit;
   by sim;
run;

/*DELETING IMPUTED VALUES AFTER THE EVENT – otherwise 'JM' Package in R cannot be applied*/
 data multipleFCSCMAR;
set mult_fcs_allomitfin_CMAR;

   array lh{*} lh1-lh7;
   if ntime_disc<=6 then do;
     if ntime_disc=2 then do;
        if ntime>1.5 then do; lh[3]=.; lh[4]=.; lh[5]=.;lh[6]=.; lh[7]=.;end;
        else do;  lh[3]=.; lh[2]=.; lh[3]=.; lh[4]=.;lh[5]=.; lh[6]=.; lh[7]=.; end;
      end;
       else if ntime_disc=3 then do;
         if ntime>2.5 then do;  lh[4]=.;lh[5]=.; lh[6]=.;lh[7]=.; end;
           else do; lh[3]=.; lh[4]=.; lh[5]=.;lh[6]=.; lh[7]=.;end;
         end;
```
27

```
      else if ntime_disc=4 then do;
         if ntime>3.5 then do;  lh[5]=.;lh[6]=.;lh[7]=.; end;
            else do; lh[4]=.;lh[5]=.;lh[6]=.;lh[7]=.;  end;
      end;
      else if ntime_disc=5 then do;
         if ntime>4.5 then do; lh[6]=.; lh[7]=.; end;
         else do;   lh[5]=.; lh[6]=.; lh[7]=.;  end;
      end;

   else if ntime_disc=6 then do;
       if ntime>5.5 then do;  lh[7]=.; end;
        else do;  lh[6]=.;lh[7]=.;  end;
      end;
   end;
run;
```

/*A MACRO FOR PREPARING 5 COMPLETED DATA SETS (MULTIPLES) OF LONGITUDINAL PROCESS for 'JM' Package in R*/

```
%macro prepmult;
  %do m=1 %to 5;

 data multiple&m.FCSALLCMAR;
   set multipleFCSCMAR;
if _imputation_=&m;
 keep sim subj female old lH1 lH2 lH3 lH4 lH5 lH6;
run;

data multiple&m.FCSALLCMAR;
 set multiple&m.FCSALLCMAR;
```

```sas
 array lhb_ar{6} lH1 lH2 lH3 lH4 lH5 lH6;
do Time=1 to 6;
     lhb=lhb_ar{time};
      output;
   end;

keep  sim subj female old time lhb;
run;

data multiple&m.FCSALLCMAR;
 set multiple&m.FCSALLCMAR;
   time=time-0.5;
run;

data one.mult&m.FCSALL_omitfinCMAR;
  set multiple&m.FCSALLCMAR;
run;
    %end;
%mend;
 %prepmult;
```

**Table 1. Results of 400 simulations (n=4,000):**

**Association between cancer risk and HbA1c level in the previous time-period, true value of logHR = 1.4**

| Missing pattern | % missing HbA1c | % "all-missing"[b] | Method | Mean logHR (95% CI) | Percent Bias | Root Mean Squared Error | Coverage[e] | $\lambda$[f]±SD |
|---|---|---|---|---|---|---|---|---|
| Fully Observed | 0% | | *Standard JM* | 1.31 (1.29, 1.32) | -7% | 0.17 | 92% | |
| Completely MAR Ratio[a] = 1.0 | 40% | 4.9 | *Standard JM*[c] | 1.31 (1.30, 1.33) | -6% | 0.17 | 92% | |
| | | | Classic FCS + JM | 1.27 (1.26, 1.29) | -9% | 0.19 | 88% | 0.08 (0.05) |
| | | | Modified FCS[d] + JM | 1.28 (1.27, 1.29) | -9% | 0.19 | 88% | 0.08 (0.05) |
| Weak Non-MAR Ratio[a] = 1.06 | 40% Overall | 5.5 | *Standard JM*[c] | 1.19 (1.17, 1.20) | -15% | 0.26 | 74% | |
| | | | Classic FCS + JM | 1.27 (1.26, 1.29) | -9% | 0.19 | 88% | 0.08 (0.05) |
| | | | Modified FCS[d] + JM | 1.29 (1.28, 1.30) | -8% | 0.18 | 90% | 0.09 (0.05) |
| Strong Non-MAR Ratio[a] = 1.26 | 40% Overall | 24.0 | *Standard JM*[c] | 0.94 (0.92, 0.96) | -33% | 0.50 | 43% | |
| | | | Classic FCS + JM | 1.03 (1.02, 1.05) | -26% | 0.41 | 58% | 0.25 (0.13) |
| | | | Modified FCS[d] + JM | 1.29 (1.28, 1.31) | -8% | 0.20 | 95% | 0.29 (0.15) |

[a]ratio of the mean of observed values to the mean of missing values in each time-period. [b]relative size of the "all-missing" subgroup, averaged on 400 simulations. [c]the "all-missing" subgroup is excluded. [d]the imputation model includes an indicator for the "all-missing" subgroup. [e]coverage calculated based on t distribution with appropriate degrees of freedom. [f]proportion of variance due to missingness.

**Table 2. Results of 400 simulations (n=4,000[a]):**

**FCS completed values (following imputation) versus fully-observed (before inducing missing values) in the "all-missing" subgroup and in the rest of the sample by type of missing and time-period[b]**

| Missing pattern | Time | "All-missing" subgroup[c] | | | | The rest of the sample | | | |
|---|---|---|---|---|---|---|---|---|---|
| | | Fully observed | Imputed by standard FCS | Imputed by modified FCS | Ratio modified to standard FCS | Fully observed | Imputed by standard FCS | Imputed by modified FCS | Ratio modified standard FCS |
| Completely MAR  Size of the "all-missing" group: N = 175[f] | 1 | 1.981[d] | 1.98[e] | 1.982[e] | 1.001 | 1.957 | 1.957 | 1.957 | 1.000 |
| | 2 | 1.964 | 1.962 | 1.965 | 1.002 | 1.939 | 1.939 | 1.939 | 1.000 |
| | 3 | 1.946 | 1.945 | 1.947 | 1.001 | 1.921 | 1.921 | 1.921 | 1.000 |
| | 4 | 1.93 | 1.929 | 1.931 | 1.001 | 1.902 | 1.902 | 1.902 | 1.000 |
| | 5 | 1.912 | 1.912 | 1.913 | 1.001 | 1.884 | 1.884 | 1.884 | 1.000 |
| | 6 | 1.895 | 1.895 | 1.896 | 1.001 | 1.866 | 1.866 | 1.866 | 1.000 |
| | 7 | 1.878 | 1.878 | 1.878 | 1.000 | 1.848 | 1.848 | 1.848 | 1.000 |
| Weak NMAR  Size of the "all-missing" group: N = 194[f] | 1 | 2.037 | 2.013 | 2.019 | 1.003 | 1.954 | 1.952 | 1.952 | 1.000 |
| | 2 | 2.021 | 1.999 | 2.007 | 1.004 | 1.936 | 1.934 | 1.934 | 1.000 |
| | 3 | 2.006 | 1.988 | 1.992 | 1.002 | 1.917 | 1.916 | 1.915 | 0.999 |
| | 4 | 1.991 | 1.972 | 1.979 | 1.004 | 1.899 | 1.897 | 1.897 | 1.000 |
| | 5 | 1.975 | 1.958 | 1.963 | 1.003 | 1.88 | 1.879 | 1.879 | 1.000 |
| | 6 | 1.959 | 1.944 | 1.949 | 1.003 | 1.862 | 1.861 | 1.861 | 1.000 |
| | 7 | 1.943 | 1.929 | 1.934 | 1.003 | 1.844 | 1.843 | 1.843 | 1.000 |
| Strong NMAR  Size of the "all-missing" group: N = 847[f] | 1 | 2.146 | 1.911 | 1.984 | 1.038 | 1.900 | 1.888 | 1.888 | 1.000 |
| | 2 | 2.131 | 1.897 | 1.989 | 1.048 | 1.88 | 1.872 | 1.871 | 0.999 |
| | 3 | 2.115 | 1.883 | 1.978 | 1.050 | 1.861 | 1.854 | 1.854 | 1.000 |
| | 4 | 2.1 | 1.868 | 1.964 | 1.051 | 1.842 | 1.837 | 1.836 | 0.999 |
| | 5 | 2.084 | 1.853 | 1.95 | 1.052 | 1.823 | 1.820 | 1.819 | 0.999 |
| | 6 | 2.068 | 1.839 | 1.938 | 1.054 | 1.804 | 1.803 | 1.802 | 0.999 |
| | 7 | 2.053 | 1.824 | 1.923 | 1.054 | 1.785 | 1.787 | 1.786 | 0.999 |

[a] after deleting events occurring on the 1st time-period, N=~3534. [b] disregarding the event. [c] the size of this subgroup is different for each missing pattern. [d] log Hba1C values averaged on 400 simulations. [e] log Hba1C values averaged on 400 simulations and 5 multiples. [f] averaged on 400 simulations.

**Table 3. Results of 1,600 simulations (n=4,000): Type-1 error**

**Association between cancer risk and HbA1c level in the previous time-period, true value of logHR = 0**

| Missing pattern | % missing HbA1c | Method | Mean logHR (95% CI) | Type-1 Error[d] |
|---|---|---|---|---|
| Weak Non-MAR Ratio[a] = 1.06 | 40% Overall | *Standard JM*[b] | -0.09 (-0.11, -0.07) | 6% |
| | | Standard FCS + JM | 0.04 (0.02, 0.06) | 5% |
| | | Modified FCS[c] + JM | 0.04 (0.02, 0.06) | 5% |
| Strong Non-MAR Ratio[a] = 1.26 | 40% Overall | *Standard JM*[b] | -0.29 (-0.30, -0.27) | 9% |
| | | Standard FCS + JM | 0.07 (0.04, 0.09) | 3% |
| | | Modified FCS[c] + JM | 0.13 (0.10, 0.15) | 4% |

[a] ratio of the mean of observed values to the mean of missing values in each time-period.

[b] the "all-missing" subgroup is excluded. [c] the imputation model includes an indicator for the "all-missing" subgroup

[d] 100 minus the coverage; coverage calculated based on t distribution with appropriate degrees of freedom

**Table 4. Baseline Characteristics of all "Clalit" insurants who were cancer-free in 2002, and diagnosed with diabetes in 2002-2012. Follow up started two years after diabetes diagnosis and ended in prostate cancer diagnosis, reaching age 90, death, or 30.12.2012**

| Characteristic | Total N=152,803 | "All-missing" subgroup N=33,420 | Rest of the group N=119,383 | p-value |
|---|---|---|---|---|
| Sex (% males) | 47 | 40 | 49 | <.0001 |
| Age at diabetes diagnosis (years), mean (SD) | 58.7 (14.7) | 56.8 (19.1) | 59.3 (13.2) | <.0001 |
| Ethnic origin, % | | | | <.0001 |
|     Ashkenazi Jews | 30.9 | 33.8 | 30.0 | |
|     Sephardic Jews | 27.5 | 27.3 | 27.5 | |
|     Israeli Born Jews | 18.1 | 17.8 | 18.1 | |
|     Israeli Arabs | 18.3 | 16.0 | 18.9 | |
|     Yemenite, Ethiopian and Central African | 5.3 | 5.1 | 5.4 | |
| Socioeconomic status (SES), % | | | | <.0001 |
|     Low | 43.9 | 43.8 | 44.0 | |
|     Medium | 37.5 | 36.4 | 37.8 | |
|     High | 15.9 | 17.1 | 15.6 | |
|     Missing | 2.7 | 2.8 | 2.7 | |
| Pancreatic cancer during follow-up (%) | 0.18 | 0.16 | 0.18 | 0.39 |
|     Died during follow-up (%) | 10.4 | 17.3 | 8.49 | <.0001 |
|     Reached age 90 during follow-up (%) | 2.2 | 4.1 | 1.7 | <.0001 |
| Years of follow-up | | | | |

| | | 3.6 (2.6) | 5.0 (2.5) | |
|---|---|---|---|---|
| Mean (sd) | | 3.6 (2.6) | 5.0 (2.5) | |
| Median | | 3.0 | 4.9 | <.0001 |

**Table 5. Motivating Example: association between pancreatic cancer and HbA1c in the previous time-period**

| Method | # events/N | Mean logHR (SE) | HR[a] (95%CI) | $\lambda$ [d] |
|---|---|---|---|---|
| *Standard JM*[b] | 220/119,328 | 3.605 (0.48) | 1.41 (1.29, 1.54) | |
| standard FCS + JM | 274/152,710 | 3.409 (0.47) | 1.38 (1.27, 1.51) | 0.20 |
| Modified FCS[c] + JM | 274/152,710 | 3.346 (0.49) | 1.37 (1.25, 1.51) | 0.26 |

[a]for an increase of 10% in HbA1c level. [b]the "all-missing" subgroup is excluded. [c]the imputation model includes an indicator for the "all-missing" subgroup. [d]proportion of variance due to missingness

**Table 6. FCS-Imputed values in the "all-missing" subgroup (N = 33,382, 22% of the sample) and in the rest of the sample. Completed marker values (average over 10 multiples) in each time-period of follow-up[a]**

|  | "All-missing" subgroup[a] | | | The rest of the sample | | |
|---|---|---|---|---|---|---|
| Time | Imputed by modified FCS | Imputed by standard FCS | Ratio between versions | Imputed by modified FCS | Imputed by standard FCS | Ratio between versions |
| 1 | 1.883[b] | 1.9[b] | 0.991 | 1.904 | 1.904 | 1.000 |
| 2 | 1.888 | 1.899 | 0.994 | 1.903 | 1.903 | 1.000 |
| 3 | 1.892 | 1.904 | 0.994 | 1.909 | 1.909 | 1.000 |
| 4 | 1.894 | 1.909 | 0.992 | 1.913 | 1.913 | 1.000 |
| 5 | 1.884 | 1.913 | 0.985 | 1.917 | 1.917 | 1.000 |
| 6 | 1.882 | 1.916 | 0.982 | 1.92 | 1.919 | 1.001 |
| 7 | 1.893 | 1.923 | 0.984 | 1.927 | 1.927 | 1.000 |
| 8 | 1.854 | 1.928 | 0.962 | 1.932 | 1.931 | 1.001 |
| 9 | 1.879 | 1.938 | 0.970 | 1.942 | 1.942 | 1.000 |
| 10 | 1.873 | 1.941 | 0.965 | 1.945 | 1.945 | 1.000 |
| 11 | 1.875 | 1.948 | 0.963 | 1.952 | 1.951 | 1.001 |
| 12 | 1.849 | 1.949 | 0.949 | 1.953 | 1.953 | 1.000 |
| 13 | 1.879 | 1.953 | 0.962 | 1.957 | 1.957 | 1.000 |
| 14 | 1.829 | 1.956 | 0.935 | 1.959 | 1.959 | 1.000 |
| 15 | 1.836 | 1.965 | 0.934 | 1.968 | 1.968 | 1.000 |
| 16 | 1.856 | 1.964 | 0.945 | 1.968 | 1.967 | 1.001 |
| 17 | 1.801 | 1.966 | 0.916 | 1.969 | 1.968 | 1.001 |
| 18 | 1.864 | 1.954 | 0.954 | 1.958 | 1.957 | 1.001 |

[a]disregarding the event. [b]log Hba1C values averaged on 10 multiples

**Figure 1: The Idea of Joint Models for a specific subject [2]**

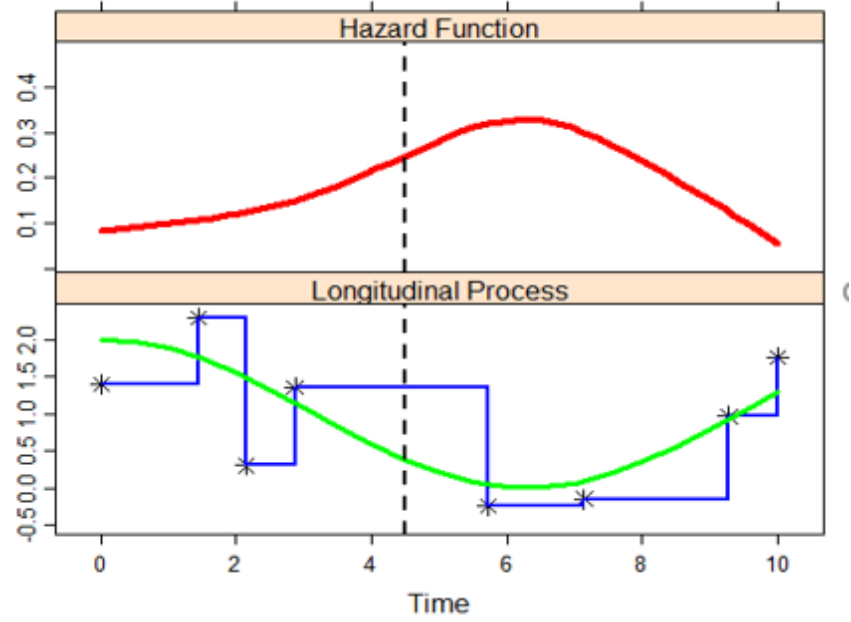

**Figure 2: Spaghetti plot of Longitudinal observed marker values of 10 randomly selected subjects (before inducing missing values)**

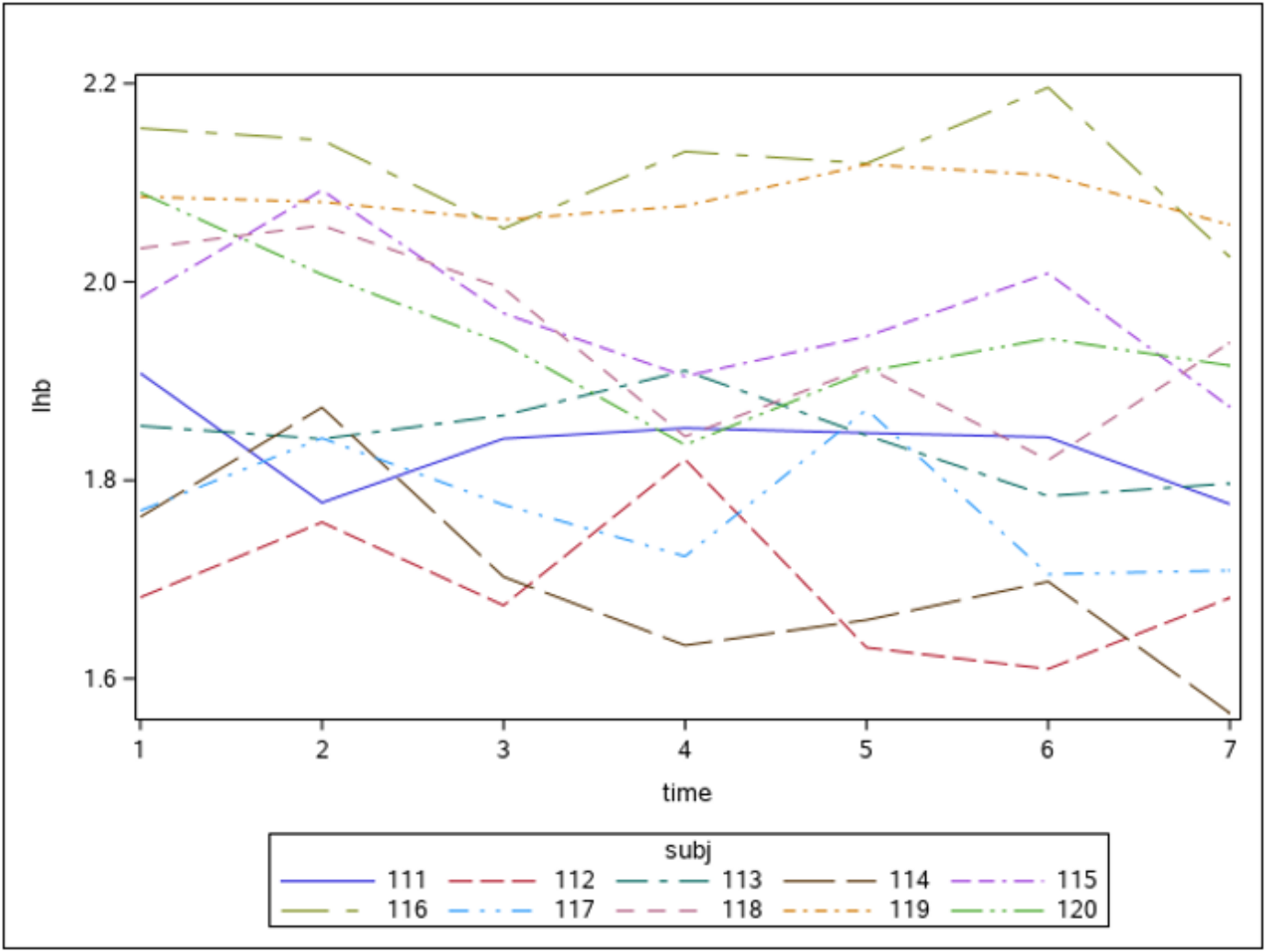